\documentstyle[12pt]{article}
\advance \topmargin by -\headheight
\advance \topmargin by -\headsep

\evensidemargin \oddsidemargin
\begin{document}

\topmargin -.6in

\def\rf#1{(\ref{eq:#1})}
\def\lab#1{\label{eq:#1}}
\def\nonu{\nonumber}
\def\br{\begin{eqnarray}}
\def\er{\end{eqnarray}}
\def\be{\begin{equation}}
\def\ee{\end{equation}}
\def\eq{\!\!\!\! &=& \!\!\!\! }
\def\foot#1{\footnotemark\footnotetext{#1}}
\def\lb{\lbrack}
\def\rb{\rbrack}
\def\llangle{\left\langle}
\def\rrangle{\right\rangle}
\def\blangle{\Bigl\langle}
\def\brangle{\Bigr\rangle}
\def\llbrack{\left\lbrack}
\def\rrbrack{\right\rbrack}
\def\lcurl{\left\{}
\def\rcurl{\right\}}
\def\({\left(}
\def\){\right)}
\newcommand{\nit}{\noindent}
\newcommand{\ct}[1]{\cite{#1}}
\newcommand{\bi}[1]{\bibitem{#1}}
\def\lskip{\vskip\baselineskip\vskip-\parskip\noindent}
\relax

\def\tr{\mathop{\rm tr}}
\def\Tr{\mathop{\rm Tr}}
\def\v{\vert}
\def\bv{\bigm\vert}
\def\Bgv{\;\Bigg\vert}
\def\bgv{\bigg\vert}
\newcommand\partder[2]{{{\partial {#1}}\over{\partial {#2}}}}
\newcommand\funcder[2]{{{\delta {#1}}\over{\delta {#2}}}}
\newcommand\Bil[2]{\Bigl\langle {#1} \Bigg\vert {#2} \Bigr\rangle}  
\newcommand\bil[2]{\left\langle {#1} \bigg\vert {#2} \right\rangle} 
\newcommand\me[2]{\left\langle {#1}\right|\left. {#2} \right\rangle}  
\newcommand\sbr[2]{\left\lbrack\,{#1}\, ,\,{#2}\,\right\rbrack}
\newcommand\pbr[2]{\{\,{#1}\, ,\,{#2}\,\}}
\newcommand\pbbr[2]{\lcurl\,{#1}\, ,\,{#2}\,\rcurl}
%
\def\a{\alpha}
\def\b{\beta}
\def\dc{{\cal D}}
\def\d{\delta}
\def\D{\Delta}
\def\eps{\epsilon}
\def\vareps{\varepsilon}
\def\g{\gamma}
\def\G{\Gamma}
\def\grad{\nabla}
\def\h{{1\over 2}}
\def\l{\lambda}
\def\L{\Lambda}
\def\m{\mu}
\def\n{\nu}
\def\o{\over}
\def\om{\omega}
\def\O{\Omega}
\def\p{\phi}
\def\P{\Phi}
\def\pa{\partial}
\def\pr{\prime}
\def\ra{\rightarrow}
\def\s{\sigma}
\def\S{\Sigma}
\def\t{\tau}
\def\th{\theta}
\def\Th{\Theta}
\def\ti{\tilde}
\def\wti{\widetilde}
\def\sj{{\jmath}{}}
\def\bsj{{\bar \jmath}{}}
\def\bp{{\bar \p}}
\def\faa{Fa\'a di Bruno~}
\newcommand\sumi[1]{\sum_{#1}^{\infty}}   
\newcommand\fourmat[4]{\left(\begin{array}{cc}  
{#1} & {#2} \\ {#3} & {#4} \end{array} \right)}
\newcommand\twocol[2]{\left(\begin{array}{c}  
{#1} \\ {#2} \end{array} \right)}
%
\def\lie{{\cal G}}
\def\alie{{\hat {\cal G}}}
\def\CKP{{\sf CKP}}
\def\GNLS{{\sf GNLS}}
\def\winf{{\sf w_\infty}}
\def\win1{{\sf w_{1+\infty}}}
\def\hwinf{{\sf {\hat w}_{\infty}}}
\def\Winf{{\sf W_\infty}}
\def\Win1{{\sf W_{1+\infty}}}
\def\hWinf{{\sf {\hat W}_{\infty}}}
\def\KP{${\sf \, KP\,}$}                 
%
\def\rlx{\relax\leavevmode}
\def\inbar{\vrule height1.5ex width.4pt depth0pt}
\def\IZ{\rlx\hbox{\sf Z\kern-.4em Z}}
\def\IR{\rlx\hbox{\rm I\kern-.18em R}}
\def\IC{\rlx\hbox{\,$\inbar\kern-.3em{\rm C}$}}
\def\one{\hbox{{1}\kern-.25em\hbox{l}}}
%
\def\mark{\noindent{\bf Remark.}\quad}
\def\prop{\noindent{\bf Proposition.}\quad}
\def\theor{\noindent{\bf Theorem.}\quad}
\def\name{\noindent{\bf Definition.}\quad}
\def\exam{\noindent{\bf Example.}\quad}
\def\proof{\noindent{\bf Proof.}\quad}
\newcommand\PRL[3]{{\sl Phys. Rev. Lett.} {\bf#1} (#2) #3}
\newcommand\NPB[3]{{\sl Nucl. Phys.} {\bf B#1} (#2) #3}
\newcommand\NPBFS[4]{{\sl Nucl. Phys.} {\bf B#2} [FS#1] (#3) #4}
\newcommand\CMP[3]{{\sl Commun. Math. Phys.} {\bf #1} (#2) #3}
\newcommand\PRD[3]{{\sl Phys. Rev.} {\bf D#1} (#2) #3}
\newcommand\PLA[3]{{\sl Phys. Lett.} {\bf #1A} (#2) #3}
\newcommand\PLB[3]{{\sl Phys. Lett.} {\bf #1B} (#2) #3}
\newcommand\JMP[3]{{\sl J. Math. Phys.} {\bf #1} (#2) #3}
\newcommand\PTP[3]{{\sl Prog. Theor. Phys.} {\bf #1} (#2) #3}
\newcommand\SPTP[3]{{\sl Suppl. Prog. Theor. Phys.} {\bf #1} (#2) #3}
\newcommand\AoP[3]{{\sl Ann. of Phys.} {\bf #1} (#2) #3}
\newcommand\PNAS[3]{{\sl Proc. Natl. Acad. Sci. USA} {\bf #1} (#2) #3}
\newcommand\RMP[3]{{\sl Rev. Mod. Phys.} {\bf #1} (#2) #3}
\newcommand\PR[3]{{\sl Phys. Reports} {\bf #1} (#2) #3}
\newcommand\AoM[3]{{\sl Ann. of Math.} {\bf #1} (#2) #3}
\newcommand\UMN[3]{{\sl Usp. Mat. Nauk} {\bf #1} (#2) #3}
\newcommand\FAP[3]{{\sl Funkt. Anal. Prilozheniya} {\bf #1} (#2) #3}
\newcommand\FAaIA[3]{{\sl Functional Analysis and Its Application} {\bf #1}
(#2) #3}
\newcommand\BAMS[3]{{\sl Bull. Am. Math. Soc.} {\bf #1} (#2) #3}
\newcommand\TAMS[3]{{\sl Trans. Am. Math. Soc.} {\bf #1} (#2) #3}
\newcommand\InvM[3]{{\sl Invent. Math.} {\bf #1} (#2) #3}
\newcommand\LMP[3]{{\sl Letters in Math. Phys.} {\bf #1} (#2) #3}
\newcommand\IJMPA[3]{{\sl Int. J. Mod. Phys.} {\bf A#1} (#2) #3}
\newcommand\AdM[3]{{\sl Advances in Math.} {\bf #1} (#2) #3}
\newcommand\RMaP[3]{{\sl Reports on Math. Phys.} {\bf #1} (#2) #3}
\newcommand\IJM[3]{{\sl Ill. J. Math.} {\bf #1} (#2) #3}
\newcommand\APP[3]{{\sl Acta Phys. Polon.} {\bf #1} (#2) #3}
\newcommand\TMP[3]{{\sl Theor. Mat. Phys.} {\bf #1} (#2) #3}
\newcommand\JPA[3]{{\sl J. Physics} {\bf A#1} (#2) #3}
\newcommand\JSM[3]{{\sl J. Soviet Math.} {\bf #1} (#2) #3}
\newcommand\MPLA[3]{{\sl Mod. Phys. Lett.} {\bf A#1} (#2) #3}
\newcommand\JETP[3]{{\sl Sov. Phys. JETP} {\bf #1} (#2) #3}
\newcommand\JETPL[3]{{\sl  Sov. Phys. JETP Lett.} {\bf #1} (#2) #3}
\newcommand\PHSA[3]{{\sl Physica} {\bf A#1} (#2) #3}
\newcommand\PHSD[3]{{\sl Physica} {\bf D#1} (#2) #3}
\newcommand{\sect}[1]{\setcounter{equation}{0}\section{#1}}
\renewcommand{\theequation}{\thesection.\arabic{equation}}
\relax
\def\cA{{\cal A}}
\def\cB{{\cal B}}
\def\cC{{\cal C}}
\def\cD{{\cal D}}
\def\cE{{\cal E}}
\def\cF{{\cal F}}
\def\cH{{\cal H}}
\def\cK{{\cal K}}
\def\cL{{\cal L}}
\def\cM{{\cal M}}
\def\cR{{\cal R}}
\begin{titlepage}
\vspace{-1cm}
\noindent
July, 1994 \hfill{IFT-P/029/94}\\
\phantom{bla}
\hfill{UICHEP-TH/93-10}\\
\phantom{bla}
\hfill{hep-th/9408104}
\\
\vskip .3in

\begin{center}

{\large\bf Affine Lie Algebraic Origin of}
\end{center}
\begin{center}
{\large\bf Constrained KP Hierarchies}
\end{center}
\normalsize
\vskip .4in

\begin{center}
{ H. Aratyn\footnotemark
\footnotetext{Work supported in part by U.S. Department of Energy,
contract DE-FG02-84ER40173 and by NSF, grant no. INT-9015799}}

\par \vskip .1in \noindent
Department of Physics \\
University of Illinois at Chicago\\
845 W. Taylor St.\\
Chicago, Illinois 60607-7059\\
\par \vskip .3in

\end{center}

\begin{center}
{J.F. Gomes\footnotemark
\footnotetext{Work supported in part by CNPq}} and A.H. Zimerman$^{\,2}$

\par \vskip .1in \noindent
Instituto de F\'{\i}sica Te\'{o}rica-UNESP\\
Rua Pamplona 145\\
01405-900 S\~{a}o Paulo, Brazil
\par \vskip .3in

\end{center}

\begin{center}
{\large {\bf ABSTRACT}}\\
\end{center}
\par \vskip .3in \noindent

We present an affine $sl (n+1)$ algebraic construction of the basic
constrained KP hierarchy.

This hierarchy is analyzed using two approaches, namely linear matrix
eigenvalue problem on hermitian symmetric space and constrained KP Lax
formulation and we show that these approaches are equivalent.

The model is recognized to be the generalized non-linear Schr\"{o}dinger
($\GNLS$) hierarchy and it is used as a building block for a new class of
constrained KP hierarchies.
These constrained KP hierarchies are connected via similarity-B\"{a}cklund
transformations and interpolate between $\GNLS$ and multi-boson KP-Toda
hierarchies.
Our construction uncovers origin of the Toda lattice structure behind
the latter hierarchy.
\end{titlepage}

\sect{Introduction}

This paper examines a class of the constrained KP hierarchies and their
mutual relations, both from the matrix and pseudo-differential
point of view.
The constrained KP hierarchy structure we uncover can be denoted
(for $n$ pairs of bosonic
fields) by a symbol $\CKP_n^{(k)}$ with index $k$ ranging from $0$ to $n$
and is associated with the Lax operators
\br
&&L_n^{(k)} \equiv \pa + a^{(k)}_1 \( \pa - S^{(k)}_k \)^{-1}
+ a^{(k)}_2 \( \pa - S^{(k-1)}_{k-1} \)^{-1} \( \pa - S^{(k)}_k \)^{-1}
+ \ldots \lab{intro1}\\
&&+ a^{(k)}_k \( \pa - S^{(1)}_{1} \)^{-1} \cdots
\( \pa - S^{(k)}_k \)^{-1} + \sum_{i=k+1}^n a^{(k)}_i
\( \pa - S_i \)^{-1} \( \pa - S^{(1)}_1 \)^{-1} \cdots
\( \pa - S^{(k)}_k \)^{-1}
\nonu
\er
all satisfying Sato's KP evolution equations.
The basic model is the one corresponding to $k=0$ with the Lax operator
\ct{oevels}:
\be
L_n \equiv L^{(0)}_n = \pa + \sum_{i=1}^n a_i \(\pa - S_i \)^{-1}
\lab{intro2}
\ee
and the remaining models with $k \geq 1$ can be derived from it via the
similarity gauge transformations at the Lax operator level.
Because of the form of the Lax operator \rf{intro2} and its abelian first
bracket structure we call this $\CKP^{(0)}_n$ model
the n-generalized two-boson KP hierarchy.
We identify it, at the matrix hierarchy level, with
the $sl (n+1)$ generalized non-linear Schr\"{o}dinger (GNLS) hierarchy
\ct{FK83}. Therefore this model can also be denoted as $\GNLS_n$.
We discuss $\GNLS_n$ model using the structure of the underlying
hermitian symmetric space and derive its recursion operator as well as
its Hamiltonian structure.
We also provide an affine Lie algebraic foundation for $\GNLS_n$
hierarchy by fitting it into the AKS ${\hat s}l (n+1)$ framework \ct{aks},
generalizing the work of \ct{FNR83} for the AKNS model.
In view of the similarity relations connecting this basic $\CKP^{(0)}_n$
model and all the other constrained KP models our construction uncovers
the $sl (n+1)$ structure behind all the $\CKP^{(k)}_n$ hierarchies.

The other extreme of the sequence $\CKP^{(k)}_n$ is the important
case of $k=n$ in which we recognize the n-boson KP-Toda hierarchy
\ct{BX9305,ANPV}, which recently appeared in a study of Toda hierarchies
and matrix models \ct{BX9212,ANPZ}.
For one bose pair with $n=1$ there is only one model, namely the
two-boson KP hierarchy \ct{BAK85,2boson} or $\CKP^{(0)}_1$,
equivalent to the AKNS model.

In section 2 we recapitulate the Zakharov-Shabat-AKNS (ZS-AKNS) scheme in the
framework of the hermitian symmetric spaces.
We work with the algebra $\lie$ containing an element
$E =  2 \mu_n \cdot H /  \a_n^2 $, which allows the decomposition
$\lie = {\rm Ker} ( {\rm ad} E) \oplus {\rm Im } ( {\rm ad} E)$.

We derive a general formula for the recursion and the
Hamiltonian operators for a particular case where the linear spectral
problem $\l E + A^{0} \Psi = \pa \Psi$ defining ZS-AKNS scheme
involves a matrix $A^{0} \in {\rm Ker} ( {\rm ad} E)$.
For $sl (n+1)$ $A^{0}$ can be written in the matrix form as
\be
A^{0} = \left(\begin{array}{ccccc}
0 &\cdots &0 &\cdots & q_1 \\
0 &0 &\cdots &0 &q_2 \\
0& & \ddots & & \vdots \\
\vdots& &  &\ddots & q_n \\
r_1 &r_2  &\cdots & r_n & 0
\end{array} \right)
\lab{1.12}
\ee
In this case we deal with the $\GNLS_n$ hierarchy.

We also analyze the ZS-AKNS formalism in case of $sl (n+1)$ algebra
connected with the linear spectral problem with $ A^{0}$:
\be
A^{0} = \left(\begin{array}{ccccc}
0 &q_1 &0 &\ldots & 0 \\
0 &0 &q_2 &\ldots &0 \\
0 & 0 &0 &\ddots  & \vdots \\
\vdots& & & \ddots & q_n \\
r_1 &r_2  &\cdots & r_n & 0
\end{array} \right)
\lab{1.26}
\ee
which belongs this time to ${\rm Ker} ( {\rm ad} E) \oplus
{\rm Im } ( {\rm ad} E)$.
We give a prescription how to obtain the recursion operator in this case,
shown in Appendix B (for $n=2$) to be equivalent to the multi-boson KP-Toda
hierarchy recursion operator.

In section 3 we derive the equations of motion for the $\GNLS_n$ model
using AKS approach \ct{aks}.
This provides the Lie algebraic foundation for this
model, which will turn out to be a basic model of the constrained
KP hierarchy.
In section 4 we prove equivalence of ZS-AKNS matrix and KP Lax formulations
for the $\GNLS_n$ hierarchy.
Our proof shows that respective recursion operators coincide leading to
identical flow equations for both hierarchies.

In section 5 we construct the general $\CKP^{(k)}_n$ models in terms of the
$\GNLS_n$ hierarchy.
We base the construction on the similarity-B\"{a}cklund transformation.
The structure of this transformation ensures that all encountered models
satisfy Sato's KP flow equation.
In particular we obtain a proof for the multi-boson KP-Toda model being the
constrained KP hierarchy and recover its discrete Toda structure \ct{similar}.

In Appendix A we collect few Lie algebraic definitions and properties.

\sect{ZS-AKNS Scheme and Recursion Operators}

Consider the linear matrix problem:
\br
A \Psi \eq \pa \Psi \lab{1.1} \\
B_m \Psi \eq \pa_{t_m} \Psi \qquad m=2,3,\ldots \lab{1.2}
\er
for
\be
A = \l E + A^{0} \qquad {\rm with} \qquad
E = { 2 \mu_a \cdot H \o \a_a^2}
\lab{1.3}
\ee
where $\mu_a$ is a fundamental weight and $\a_a$ are
simple roots of $\lie$.
The element $E$ is used to decompose the Lie algebra
$\lie$ as follows:
\be
\lie = \cK \oplus \cM = {\rm Ker} ( {\rm ad} E) \oplus
{\rm Im } ( {\rm ad} E)
\lab{1.5}
\ee
where the $\cK \equiv {\rm Ker} ( {\rm ad} E)$  has the form $
\cK ^{\pr} \times u(1)$ and is spanned by
 the Cartan subalgebra of $\lie$ and step operators associated to roots not
containing $\a_a$.
Moreover $\cM \equiv {\rm Im } ( {\rm ad} E)$ is the orthogonal complement of
$\cK$.

In fact, if ${2\mu_a .\a /{\a_a^2}} = \pm 1 ,0$,  the above
 decomposition defines an hermitian symmetric space
$ \lie / \cK $ \ct{FK83}.
This special choice of $E$ will play a crucial role in what
follows.

The compatibility condition for the linear problem \rf{1.1} leads to the
Zakharov-Shabat (Z-S) integrability equations
\be
\pa_m A - \pa B_m + \lb A \, , \, B_m \rb = 0 \quad;\quad
\pa \equiv \pa_x \quad \pa_{t_m} \equiv \pa_m
\lab{1.6}
\ee
Inserting the decomposition of $A$ from \rf{1.3} into \rf{1.6} we get
\be
\pa_m A^{0} - \pa B_m + \l \lb E \, , \, B_m \rb
+\lb A^{0} \, , \, B_m \rb = 0
\lab{1.7}
\ee
We search for solutions of \rf{1.7} of the form
\be
B_{m} = \sum_{i=0}^{m} \l^{i} B^{i}_{m}
\lab{1.8}
\ee
To determine the coefficients of the expansion in \rf{1.7} we first
find {}from \rf{1.8} the following identity:
\be
\pa_{m} A^{0} - \l \pa_{m-1} A^{0} +\pa X (\l ) +
\l \lb E \, , \, X (\l) \rb + \lb A^{0} \, , \, X (\l) \rb =0
\lab{1.9}
\ee
where $X (\l) = B_{m} - \l B_{m-1} $.
By comparing powers of $\l$ in \rf{1.9} we can split
$X (\l)$ as
\be
X (\l) = O_m  (1) + Y_m (\l) \qquad ; \qquad Y_m (\l) \in \cK
\lab{1.10}
\ee
where $O_m  (1)$ is an arbitrary element of algebra $\lie$ independent
of $\l$. Therefore, all dependence on $\l$ is contained
in $Y_m (\l)$, an element of $\cK$.

We now consider two distinct cases defined according to
whether $A^{0}$ is entirely in $\cM$ or not.

\subsection{${\bf A^{0} \in Im(ad E)}$}

Since now $ A^{0} \in \cM$ equation \rf{1.9} yields for the $\l$ dependent
element $Y_m$ in $\cK$ the condition $ \pa Y_m (\l) =0$
and hence this component of $X(\l)$ can be chosen to vanish.
With this choice, equation \rf{1.10} becomes a so-called congruence
relation:
\be
B_m = \l B_{m-1} + O_m  (1) = \sum_{i=0}^{m} O_{m-i} (1) \l^i
\lab{congru}
\ee
Plugging \rf{congru} into \rf{1.7} we find that the coefficients $O_j  (1)$
with $1 \leq j \leq m-1$ satisfy the recursion relation
\be
\pa O_j  (1) - \sbr{A^{(0)}}{O_j  (1)} = \sbr{E}{O_{j+1} (1)}  \lab{congr1}
\ee
There is a compact way to rewrite all these recurrence relations for all $m$.
It is provided by a single equation:
\be
\pa B= \sbr{\l E + A^{(0)}}{B}  \lab{novik}
\ee
where $B$ is now an infinite series:
\be
B \equiv \sum_{j=0}^{\infty} O_{j} (1) \l^{-j}
\lab{novikb}
\ee

Since $Y_m=0$, the equation \rf{1.9} splits into two simple
expressions for coefficients of $\l^i,\, i=0,1$
\br
\pa_{m-1} A^{0} &=& \lb E \, , \, O_m (1) \rb \lab{1.13} \\
\pa_{m} A^{0} &=&  \pa O_m (1) - \lb A^{0}  \, , \, O_m (1) \rb
\lab{1.14}
\er
Equation \rf{1.13} can easily be inverted
to yield expression for $O_m^{\cM} (1)$.

{}From the relation \rf{A5} the solution of \rf{1.13} is
\be
O_m^{\cM} (1)  =  \lb E \, , \, \pa_{m-1} A^{0}  \rb
\lab{1.16}
\ee
{}From \rf{1.14} we find now
\be
\pa O_m^{\cK} (1) = \lb A^{0}   \, , \, O_m^{\cM} (1) \rb
\quad \to \quad O_m^{\cK} (1) = \pa^{-1} \lb A^{0}   \, , \,
\lb E \, , \, \pa_{m-1} A^{0}  \rb  \rb
\lab{1.17}
\ee
displaying the non-local character of $O_m^{\cK} (1)$.
Plugging now these two results into \rf{1.14} we find
\be
\pa_m A^{0} = \lb E \, , \, \pa \pa_{m-1} A^{0}  \rb
- \lb A^{0} \, , \, \pa^{-1}
\lb A^{0}   \, , \, \lb E \, , \, \pa_{m-1} A^{0}  \rb  \rb  \rb
= \cR \pa_{m-1} A^{0}
\lab{1.18}
\ee
where we have defined a recursion operator \ct{FK83}
\be
\cR \equiv  \( \pa - ad_{A^0}\, \pa^{-1} ad_{A^0} \) ad_{E}
\lab{recoper}
\ee
Since $\cM$ is spanned by $2n$ step operators associated to roots containing
$\a_a$ only once, i.e. $\a = \a_a + \ldots$ we denote $E_{\pm(\a_a + \ldots)}
= E_{\pm}^a$, where $a=1, \ldots, n$.
We thus parametrize this model by
\be
A^{(0)} = \sum_{a=1}^n \( q_a E_{+}^a + r_a E_{-}^a\)
\lab{azero}
\ee
With this parametrization we have
\br
&&\pa_n \twocol{r_i}{q_l} = \cR_{(i,l),(j,m)} \twocol{r_j}{q_m} =
\lab{R}\\
&&\fourmat{-\pa + r_s \pa^{-1} q_k R^{i\,*}_{sj-k}}
{r_s \pa^{-1} r_k R^{i\,*}_{sk-m}}{-q_s \pa^{-1} q_kR^{l}_{sk-j}}
{\pa - q_s \pa^{-1} r_k R^{l}_{sm-k}}
\pa_{n-1} \twocol{r_j}{q_m}
\nonu
\er
where $R^{l}_{sm-k}$ and $R^{i\,*}$ are defined in \rf{A8}.
We now find conserved Hamiltonians and the first Hamiltonian bracket
structure associated to the Z-S equations \rf{1.6} in the Hamiltonian
form:
\be
\pa_m A^{0} = \pbr{H_m}{A^{0}}
\lab{1.20}
\ee
{\bf Lemma:} $\qquad \qquad{\bf  \pa_n \Tr \({A^{0}}^2 \)=
-2 \pa\Tr \( E O_{n+1} (1) \)}$

The proof follows from the following calculation:
\br
\pa_{n}\Tr \(A^{0}  A^{0} \)\eq
 2\Tr \( \lb E, \lb  E, A^{0}\rb \rb \pa_{n} A^{0} \)
= 2 \Tr \(E \lb A^{0} , \lb \pa_{n} A^{0}, E \rb \rb \)
\nonu \\
\eq - 2\pa \Tr \( E O_{n+1}^{\cK} (1) \)= - 2 \pa \Tr \( E O_{n+1} (1)\)
\lab{1.21}
\er
Consider now the Hamiltonian density
 $\cH_1 = \sum_{i=1}^{n}q_i r_i = \Tr \(A^{0}\)^2$ and impose
the basic relation $ \cH_n = \pa^{-1} \pa_n \cH_1$ for higher Hamiltonians.
It follows now from the above Lemma that $ \cH_{n}
 = - 2 \Tr \( E O_{n+1} (1) \)$,
which coincides with the result of Wilson \ct{wilson} (see also
\ct{crumey}).
%
The Z-S equations can be obtained from the Hamiltonians defined above
and the first bracket structure given by
\be
P_1 \d (x-y) =
\fourmat{\pbr{r_i}{r_j}}{\pbr{r_i}{q_m}}{\pbr{q_l}{r_j}}{\pbr{q_l}{q_m}}
= \fourmat{0}{-I}{I}{0} \d (x-y)
\lab{1.22}
\ee
Expressing the recursion operator $\cR$ in terms of the first
two bracket structures through
\be
\cR = P_2 P_1^{-1}
\lab{1.23}
\ee
we may obtain the closed expression for the second bracket
structure of ZS-AKNS hierarchy
\be
P_2 = \fourmat{r_s \pa^{-1} r_k R^{i\,*}_{sk-m}}
{\(\pa - r_s \pa^{-1} q_k R^{i\,*}_{sj-k} \)}
{\(\pa -q_s \pa^{-1} r_kR^{l}_{sm-k} \)}
{q_s \pa^{-1} q_kR^{l}_{sk-j}}
\lab{1.24}
\ee
This already suggests that the model is bihamiltonian.

We now consider $\lie = sl(n+1)$ with roots $\a = \a_i +
 \a_{i+1} +\ldots + \a_j $ for some $i,j=1,\ldots ,n$ , $E = {2\mu_n.H \O
{\a_n^2}}$, $\mu_n$ is the $n^{\rm th}$ fundamental weight and $H_a$, $a=1,
\ldots , n$ are the generators of the Cartan subalgebra. This
decomposition  generates the symmetric space $sl(n+1)/sl(n)\times u(1) $
(see appendix A for details).

In matrix notation we have:
\be
E = {1 \o n+1} \left(\begin{array}{ccccc}
1 & & & & \\  & 1& & & \\  & &1 & & \\ & & &\ddots &\\
& & & & -n \end{array} \right)
\lab{1.4}
\ee
This model is defined by $A^0 \in \cM$ and is parametrized with fields
$q_a$ and $r_a$, $a=1,\ldots ,n$
\be
A^{0}= \sum_{a=1}^n \( q_a E_{(\a_a+ \ldots+\a_n)}+ r_a E_{-(\a_a+ \ldots
+\a_n )}  \)
\lab{1.11}
\ee
which in the matrix form can be written as \rf{1.12}.
The successive flows \rf{1.18} related by the recursion operator \rf{R}
are given by
\br
&&\pa_n \twocol{r_i}{q_l} = \cR_{(i,l),(j,m)} \twocol{r_j}{q_m} =
\lab{1.19}\\
&&\fourmat{\(-\pa + r_k \pa^{-1} q_k\)\d_{ij}+r_i\pa^{-1}q_j}
{r_i \pa^{-1} r_m +r_m\pa^{-1}r_i}{-q_l \pa^{-1} q_j -q_j\pa^{-1}q_l}
{\(\pa -q_k\pa^{-1}r_k\)\d_{lm} -q_l \pa^{-1} r_m}
\pa_{n-1} \twocol{r_j}{q_m}
\nonu
\er
{}From \rf{1.24} an explicit expression for the second bracket is
found in this case to be:
\be
P_2 = \fourmat{r_i \pa^{-1} r_j + r_j \pa^{-1} r_i}
{\(\pa -\sum_k r_k \pa^{-1} q_k \)\d_{im} - r_i \pa^{-1} q_m}
{\(\pa -\sum_k q_k \pa^{-1} r_k \)\d_{lj} - q_l \pa^{-1} r_j}
{q_l \pa^{-1} q_m + q_m \pa^{-1}q_l}
\lab{P2}
\ee

\subsection{${\bf A^{0} \in Im(ad E)+Ker(ad E)}$}

The non-zero component $A^{0}_{\cK}$ in $A^{0}$ leads to non-zero
terms in equation \rf{1.9} in $\cK$ given by
\be
\pa_{m} A^{0}_{\cK} - \l \pa_{m-1} A^{0}_{\cK} +\pa Y_m (\l ) +
+ \lb A^{0}_{\cK}  \, , \, Y_m (\l) \rb
+ \lb A^{0}_{\cM} \, , \, O_m^{\cM} (1) \rb =0
\lab{1.27}
\ee
where $O_n^{\cM} (1)$ is the component of ${O_n (1)}$ in the subspace
$\cM$.

{}From equation \rf{1.27} we see that $ Y_m (\l) = \l Y_m \in \cK$.
The $\l$-dependent components in subspaces $\cM$ and $\cK$ read
respectively:
\br
&&-\pa_{m-1} A^{0}_{\cM} + \lb E \, , \, O_m^{\cM}(1) \rb
+ \lb A^{0}_{\cM} \, , \, Y_m \rb =0 \lab{1.28}\\
&&-\pa_{m-1} A^{0}_{\cK} - \pa Y_m +
 \lb A^{0}_{\cK} \, , \, Y_m \rb =0 \lab{1.29}
\er
while the $\l$ independent part leads to
\br
&&\pa_{m} A^{0}_{\cM} -\pa O_m^{\cM}  (1) +
 \lb A^{0}_{\cM} \, , \, O_m^{\cK}  (1) \rb
+ \lb A^{0}_{\cK} \, , \, O_m^{\cM}  (1) \rb
=0 \lab{1.30}\\
&&\pa_{m} A^{0}_{\cK} - \pa O_m^{\cK}  (1) +
 \lb A^{0}_{\cM} \, , \, O_m^{\cM}  (1) \rb
+ \lb A^{0}_{\cK} \, , \, O_m^{\cK}  (1) \rb
\lab{1.31}
\er
{}From \rf{1.29} we find
\be
Y_m = - \sum_{j=1}^{\infty } \( \pa^{-1} ad_{A^{0}_{\cK}} \)^{j-1}
\pa^{-1} \pa_{m-1} A^{0}_{\cK}
\lab{1.32}
\ee
while \rf{1.28} gives
\be
O_m^{\cM}  (1) = {\rm ad}_E \pa_{m-1} A^{0} +
{\rm ad}_E {\rm ad}_{A^{0}_{\cM}} \(
\sum_{j=1}^{\infty } \( \pa^{-1} {\rm ad}_{A^{0}_{\cK}} \)^{j-1}
\pa^{-1} \pa_{m-1} A^{0}_{\cK}\)
\lab{1.33}
\ee
It is difficult to present a closed form  solution for
$O_m^{\cK}(1)$ in general case.
However there is a simple argument for the existence of an unique
solution.
Notice that \rf{1.30} provides $n^2 - n +1$ algebraic equations (which are
not equations of motion) whilst \rf{1.31} provide $n-1$ such equations
making a total of $n^2$ algebraic equations to solve
$ n^2={\rm dim}\, \cK$ unknowns  in $O_m^{\cK}$.
Explicit examples can be worked out case by case (see Appendix B for $n=
2$).

 Let us consider a particular class of models which are connected to the
Toda Lattice hierarchy. The models are defined by a Drinfeld-Sokolov
linear system of the type
\be
\left(\begin{array}{ccccc}
\pa-\l/(n+1) &q_1 &\cdots &0 &0 \\
0 & \pa-\l/(n+1) &q_2 &\cdots &0 \\
\vdots & &\ddots & &\vdots \\ 0 & & &\pa-\l/(n+1)  & q_n\\
r_1& r_2 &\cdots &r_n & \pa+n\l/(n+1)  \end{array} \right)
\left(\begin{array}{l}
\psi_1 \\ \psi_2 \\ \vdots \\  \psi_n \\ \psi_{n+1}
\end{array} \right)   =        0
\lab{4.1}
\ee
with
\be
A^{0}= \sum_{a=1}^n q_a E_{(\a_a)}+ r_a E_{-(\a_a+ \ldots
+\a_n )}
\lab{A1.25}
\ee
which in the matrix form can be written as in \rf{1.26}.

We now argue that a more general case, with non zero upper triangular
elements may be reduced to \rf{1.26} after suitable redefinition of fields
and basis vectors.  For this purpose consider the generalized DS linear system
\be
\left(\begin{array}{ccccc}
\pa&q_1 &a_{1,3} &\cdots &a_{1,n-1} \\
0 & \pa &q_2 &\cdots &a_{2, n-1} \\
\vdots & &\ddots &q_{n-1} &a_{n, n-1} \\ 0 &0 &\cdots &\pa  & q_n\\
r_1& r_2 &\cdots &r_n & \pa+\l  \end{array} \right)
\left(\begin{array}{l}
\psi_1 \\ \psi_2 \\ \vdots \\  \psi_n \\ \psi_{n+1}
\end{array} \right)   =        0
\lab{YYYY}
\ee
{}From where we find
\be
\pa \psi_n + q_n \psi_{n+1} =0
\lab{psin}
\ee
and
\be
\pa \psi_{n-1} + q_{n-1} \psi_{n} + a_{n,n-1}\psi_{n+1} = 0
\lab{psin1}
\ee
The elimination of $\psi_{n+1}$ in favor of $\psi_n$ in \rf{psin} after
reintroducing it in \rf{psin1} we get
\be
\pa \tilde \psi_{n-1} + \tilde q_{n-1}\psi_n =0
\ee
where we have redefined
\be
\tilde \psi_{n-1}=(\psi_{n-1} - {a_{n,n-1}\o {q_n}}\psi_n)
\ee
and
\be
\tilde q_{n-1} = q_{n-1} + \pa ({a_{n,n-1} \o {q_n}})
\ee
We have therefore removed away $a_{n,n-1}$ from the linear system.
After a succession of such redefinitions, we are able to remove all
$a_{i,j}$ from the linear system showing that \rf{YYYY} can be reduced
to \rf{4.1}.
It can also be shown that the similar construction  works
in the corresponding Lax formalism.

\sect{AKS and the Generalized Non Linear Schr\"{o}dinger Equations}
We now derive the GNLS equations from the AKS formalism \ct{aks}
revealing an affine
Lie algebraic structure underlying the integrability of the hierarchy
studied  by Fordy and Kulish \ct{FK83} in connection with the symmetric
space $ sl(n+1)/sl(n)\times u(1) $.
This section provides a generalization of the work of Flaschka,
Newell and Ratiu \ct{FNR83} of the $sl(2)$ case.
The main idea lies in associating an affine Lie algebraic structure
(loop algebra $\alie \equiv \lie \otimes \IC \lb \l , \l^{-1} \rb$
of $\lie $) to an Hamiltonian structure leading to the same equations
of motion as in \rf{1.19}.
Here we only consider $\lie = sl(n+1)$ leading to $\GNLS_n$ hierarchy.
For the AKS  treatment of general symmetric spaces see \ct{harnad}.

The AKS formalism will provide us with a set of Casimir-like Hamiltonians
in involution.
The standard construction is based on the following functions
\be
\p_k (X) \equiv \h {\rm Res} \( \l^{k-1} \Tr (X^{2}) \) \qquad;\qquad X
=\sum_j X_j \l^j \in \alie
\lab{pkdef}
\ee
where $k \in \IZ_{+} $.
Define now a subalgebra $\alie^{-}_0$ of $\alie$ as $\alie^{-}_0 \equiv
\sum_{j\geq 0} \lie \otimes\l^{-j}$.
On the subalgebra $\alie^{-}_0$ there is a natural Poisson bracket:
\be
\pbr{f}{g} (X) \equiv - \bil{X}{\sbr{\pi_+ \nabla f (X)}{\pi_+ \nabla g (X)}}
\qquad;\qquad X \in \alie^{-}_0
\lab{lpbra}
\ee
where $\me{X}{Y} = \sum_{i+j=0} \Tr X_j Y_j $ and
$\pi_+$ is the canonical projection $\pi_+ : \alie \to
\alie^{+}_0 \equiv \sum_{j\geq 0} \lie \otimes\l^{j}$.

A calculation yields $ \nabla \p_k (X) = S^k X$ where $S^k$ is the shift
$S^k : X = \sum X_j \l^j \to \sum X_j \l^{j+k}$.
Consequently $\sbr{\nabla \p_k (X)}{X} = 0$, meaning that $\p_k$ is
ad-invariant.
The main AKS theorem \ct{aks} ensures now that $\p_k (X)$ restricted to
$\alie^{-}_0$ are in involution: $\pbr{\p_k}{\p_{k^{\pr}}} (X) = 0$
and generate commuting Hamiltonian flows.
For the parametrization of $\alie^{-}_0$ given by
$ \sum_{j\geq 0} X_j \l^{-j}$ with $ X_j = h_a^{(j)} H_a +
E_{\pm \a }^{(j)} E_{\pm \a}$ we obtain:
\be
\phi _k =  \sum_{i=0}^{k} {1\o {2}}g^{ab} h_a^{(i)} h_b^{(k-i)}
+ \sum_{\a >0} E_{\a }^{(i)}E_{-\a }^{(k-i)}
\lab{casimir}
\ee
where $g^{ab} = \h \a_b^{2} K^{-1}_{ab} $, $K_{ab} $ is the Cartan matrix
for $sl(n+1)$ , $E_{\pm \a } $ are step operators associated to the roots
$\pm \a $. Summation over repeated indices is understood.
The Poisson brackets \rf{lpbra} for $h_a^{(i)}$ and $E_{\a }^{(i)}$
(considered as functions on $\alie^{-}_0$) are
\br
\{ h_a^{(i)},h_b^{(j)}\} & = & 0 \nonumber \\
\{ h_a^{(i)}, E_{\a }^{(j)} \} & = & K_{\a a} E_{\a }^{(i+j)} \nonumber
\er
\be
\{ E_{\a }^{(i)}, E_{\b}^{(j)} \}  =
\left\{ \begin{array}{cl}
 \eps (\a ,\b ) E_{\a +\b}^{(i+j)} & \a +\b\quad {\rm is~a~root} \\
 \sum _{a=1}^{n} n_ah_a^{(i+j)} & \a +\b =0 \\
 0 &{\rm otherwise} \end{array} \right.
\lab{sunalgebra}
\ee
where $K_{\a a} = {2 \a . \a _a \o {\a _a^2}}$ , $\a  = \sum n_a \a _a$
and $\a _a , a=1,\ldots,n$ are the simple roots of $sl(n+1)$ \foot{
For the $sl(n+1)$ algebra the root system is given in terms of the
simple roots as $\a = \a _{a} + \a _{a+1} + \ldots + \a _{a+l}$
with $a = 1,\ldots,n$ and $l=0, 1,\ldots,n-1$
(see appendix A for explanation and notation).}.

Hamilton equations in components take the form
\be
{\pa T^{(j)} \o {\pa_{t_k}}} = \{ \phi_{k}, T^{(j)}\} \qquad;\qquad
T^{(j)} = h_a^{(j)}, E_{\a }^{(j)}
\lab{ham}
\ee
where the bracket is the Lie-Poisson bracket \rf{sunalgebra}.

The equations of motion are derived from \rf{ham} together with the
restrictions for the following leading terms
\br
h_a^{(1)} &=& 0 \nonumber \\
E_{\a}^{(0)} &=& 0, \quad {\rm for~all~roots~ } \a \nonumber \\
E_{\a_n+ \ldots+\a_a}^{(1)} & =& q_a,\quad  a=1,\ldots,n  \nonumber \\
E_{-(\a_n+\ldots+\a_a)}^{(1)} &=& r_a,\quad  a=1,\ldots,n \nonumber \\
E_{\pm (\a _a +\ldots+ \a _b )}^{(1)} & = & 0,\quad  a,b = 1,\ldots,n-1
\lab{initial}
\er
which subsequently will determine  the model.

{}From eqn \rf{ham} we find the evolution of the step
operator $E_{\b}^{(j)}$ to be
\br
{\pa E_{\b}^{(j)} \o {\pa t_k}}& =& - \h \sum _{i=0}^{k}
\sum_{a=1}^{n} n_a \(h_a^{(i)}E_{\b}^{(k+j-i)} +
h_a^{(k-i)}E_{\b}^{(i+j)} \) \nonu \\
&+& \sum _{i=0}^{k} \sum_{\a \neq \b , \a >0} \( \eps (\a,\b)E_{\a
+\b}^{(i+j)}E_{-\a}^{(k-i)} +
\eps (-\a,\b)E_{\a}^{(i)}E_{\b-\a}^{(j+k-i)} \) \nonu\\
& +&\sum _{i=0}^{k} \sum _{a=1}^{n} n_a \(\theta(\b) E_{\b}^{(i)}
h_a^{(j+k-i)} + \theta (-\b)h_a^{(i+j)}E_{\b}^{(k-i)} \)
\lab{Eevolution}
\er
for $\b = \sum n_a \a _a$ and $\theta$ the Heaviside function.
It is easy to see that $\pa E_{\b}^{(0)} /{\pa_{t_k}} = 0$
for all $k$ once we impose the condition \rf{initial} on the right hand side
of \rf{Eevolution}. This shows that $E_{\b}^{(0)}=0$ from \rf{initial}
is preserved by the flows.
If we now consider $k=j=1$ we find after
using the  conditions \rf{initial}
\be
{\pa E_{\b}^{(1)} \o {\pa t_1}} = - \sum_{a=1}^{n} n_ah_a^{(0)} E_{\b}^{(2)}
\lab{E1}
\ee
Consider now a general root not containing $\a_n$ of the form
$\b = \sum_{a=1}^{n} n_a \a_a = \a_l +\a_{l+1} + \cdots + \a_s$
with $s \leq n-1$.
Since, in this case $n_n= 0$, from the condition \rf{initial}, i.e.
$E_{\b}^{(1)} =0$ we may choose  $h_l^{(0)} =0$ for
$l=1,\ldots,n-1$ allowing therefore a nontrivial ``evolution"
for $E_{\b}^{(2)}$.
For $\b = \a_n+\ldots +\a_a$, and renaming $\pa _{t_1} = \pa _x$ we find
\be
 E_{\a_n+\ldots +\a_a}^{(2)} =\pa _xq_a
\lab{E2+}
\ee
after choosing a normalization $h_n^{(0)} = -1$. In a similar manner we have
\be
E_{-(\a_n+\ldots+\a_a)}^{(2)} = -\pa _xr_a
\lab{E2-}
\ee
Let us now consider the evolution of $h_a^{(j)}$ given from \rf{ham} by
\be
{\pa h_a^{(j)} \o {\pa t_k}} = \sum_{i=0}^k \sum _{\a >0}
K_{\a,a} \(E_{\a}^{(i)} E_{-\a}^{(j+k-i)} - E_{\a}^{i+j}E_{-\a}^{(k-i)}\)
\lab{hja}
\ee
Splitting the sum over the positive roots in terms of its simple root content,
i.e. $\sum_{\a >0} = \sum_{s=0}^{n-1} \sum_{b=1}^{n-s}$ we find
\br
{\pa h_{a}^{(j)} \o {\pa t_k}} &=&
\sum_{i=0}^k \sum_{s=0}^{n-1}\sum_{b=1}^{n-s}
\(E_{(\a_b+\ldots+\a_{b+s})}^{(i)}E_{-(\a_b+\ldots+
\a_{b+s})}^{(j+k-i)} - E_{-(\a_b+\ldots+\a_{b+s})}^{(k-i)}
E_{(\a_b+\ldots+\a_{b+s})}^{(j+i)}\) \times \nonu\\
&\times& (K_{b,a}+K_{b+1,a}+\ldots+K_{b+s,a})
\lab{hj}
\er
It is easy to verify that $\pa h_{a}^{(1)} / {\pa t_k} =0 $ for
$a=1,\ldots,n$ after imposing the condition \rf{initial} on the right
hand side of \rf{hj}. Hence the flows preserve the condition
$h^{(1)}_a= 0$ from \rf{initial}.
For $h^{(2)}_l $ we find
\be
{\pa h_{l}^{(2)} \o {\pa t_1}} = \sum
_{b=1}^{n}(q_bE_{-(\a_b+\ldots+\a_n)}^{(2)} -
r_bE_{(\a_b +\ldots+\a_n)}^{(2)})(\delta_{l,b} - \delta_{l,b-1}) =
\pa_x (q_{l+1}r_{l+1}) - \pa_x (q_lr_l)
\lab{h2}
\ee
for $l=1,\ldots,n-1$.  Due to explicit form of the Cartan matrix $K_{ab}$,
 $h_n^{(2)}$ requires an independent calculation yielding
\br
 {\pa h_{n}^{(2)} \o {\pa t_1}} &=& \sum _{l=1}^{n-1}
q_lE_{-(\a_l+\ldots+\a_n)}^{(2)} - r_lE_{(\a_l+\ldots+\a_n)}^{(2)}) +
2(q_nE_{-\a_n}^{(2)} - r_nE_{\a_n}^{(2)}) \nonumber \\
&=& -\sum_{a=1}^{n} \pa_x (q_ar_a)  - \pa_x (q_nr_n)
\lab{hn}
\er
By plugging \rf{E2+}, \rf{E2-} in \rf{hn} and choosing
the integration constants to vanish we  get
\be
h_l^{(2)} =\left\{ \begin{array}{ll}
 q_{l+1}r_{l+1} - q_lr_l & l=1,\ldots,n-1 \nonumber \\
 -\sum_{a=1}^n q_ar_a - q_nr_n & l = n  \end{array} \right.
\lab{hintegr}
\ee
We now determine the evolution of the step operators $E_{\b}^{(2)}$.
{}From \rf{Eevolution}
we find
\br
{\pa E_{\a_l+\a_{l+1}+\ldots+\a_s}^{(2)} \o {\pa t_1}} &=& \eps
(\a_{s+1}+\ldots +\a_n,\a_l +\ldots +\a_s )\,r_{s+1}\,
E_{\a_l+\ldots +\a_n}^{(2)}
\nonu\\
&+& \eps (-(\a_{l}+\ldots +\a_n),\a_l +\ldots +\a_s )
\, q_{l} \, E_{-(\a_{s+1}+\ldots +\a_n)}^{(2)}
\lab{integ}
\er
Using eqns \rf{E2+} and \rf{E2-}  and the relation \rf{A2}
implying
\be
\eps (\a_{s+1}+\ldots +\a_n,\a_l +\ldots +\a_s ) = - \eps
(-(\a_{l}+\ldots +\a_n),\a_l +\ldots +\a_s )
\lab{epsrelation}
\ee
we may integrate eqn \rf{integ} again
choosing the integration constants to vanish.
Thus we find
\be
E_{\a_l+\a_{l+1}+\ldots +\a_s}^{(2)} = \eps
(\a_{s+1}+\ldots +\a_n,\a_l +\ldots +\a_s ) \, q_l \, r_{s+1}
\lab{E+2}
\ee
In a similar manner we get
\be
E_{-(\a_l+\a_{l+1}+\ldots +\a_s)}^{(2)} = \eps
(\a_{s+1}+\ldots +\a_n,\a_l +\ldots +\a_s ) \, r_l \, q_{s+1}
\lab{E-2}
\ee
Using equation \rf{Eevolution},\rf{E2+} and \rf{E2-}
it is not difficult to see that
\be
 \pa_x^2 q_a ={\pa E_{\a_n+\a_{n-1}+\ldots +\a_a}^{(2)} \o {\pa t_1}} =
  E_{\a_n+\a_{n-1}+\ldots +\a_a}^{(3)} + 2q_a\sum_{b=1}^{n} q_br_b
\lab{evolqa}
\ee
and
\be
-  \pa_x^2 r_a ={\pa E_{-(\a_n+\a_{n-1}+\ldots +\a_a)}^{(2)} \o {\pa t_1}} =
 - E_{-(\a_n+\a_{n-1}+\ldots +\a_a)}^{(3)} - 2r_a\sum_{b=1}^{n} q_br_b
\lab{evolra}
\ee
Finally, from \rf{Eevolution} we obtain
\be
{\pa q_a \o{\pa t_2}} =
 {\pa E_{\a_a+\a_{a+1}+\ldots +\a_n}^{(1)} \o {\pa t_2}} =
  E_{\a_a+\a_{a+1}+\ldots +\a_n}^{(3)}
\lab{220}
\ee
and
\be
{\pa r_a \o {\pa t_2}} =
{\pa E_{-(\a_a+\a_{a+1}+\ldots +\a_n)}^{(1)} \o {\pa t_2}} =
-  E_{-(\a_a+\a_{a+1}+\ldots +\a_n)}^{(3)}
\lab{221}
\ee
where we made use of relations \rf{A2} and \rf{A3} between the
$\eps$'s.

Equations \rf{evolra} and \rf{evolqa} yield the following equations of motion
after inserting \rf{220} and \rf{221}
\br
 {\pa q_a \o{\pa t_2}}& =& \pa_x^2 q_a - 2q_a\sum_{b=1}^{n} \, q_b \, r_b
\nonumber \\
{\pa r_a \o{\pa t_2}} &=& -\pa_x^2 r_a + 2r_a\sum_{b=1}^{n} \,q_b \, r_b
\lab{nls}
\er
for $a=1,\ldots,n$.  These are the GNLS equations \ct{FK83}.

\sect{The n-Generalized Two-boson KP Hierarchy and its Equivalence to
${\bf sl (n+1)}$ GNLS Hierarchy}

In this section we will establish a connection between $\GNLS$
matrix hierarchy for the hermitian symmetric space $sl (n+1)$ and
 the constrained KP hierarchy.

The connection is first established between linear systems defining both
hierarchies.
Recall first the linear problem \rf{1.1}, which for $A^{0} \in
{\rm Ker} ( {\rm ad} E)$ is parametrized according to \rf{azero}:
\be
\left(\begin{array}{ccccc}
\pa-\l/(n+1) &0 &\cdots &0 &q_1 \\
0 & \pa-\l/(n+1) &0 &\cdots &q_2 \\
\vdots & &\ddots & &\vdots \\ 0 & & &\pa-\l/(n+1)  & q_n\\
r_1& r_2 &\cdots &r_n & \pa+n\l/(n+1)  \end{array} \right)
\left(\begin{array}{l}
\psi_1 \\ \psi_2 \\ \vdots \\  \psi_n \\ \psi_{n+1}
\end{array} \right)   =        0
\lab{3.1}
\ee
Perform now the phase transformation:
\be
\psi_k \; \longrightarrow \; {\bar \psi_k} = \exp \( - {1 \o n+1} \int
\l dx \) \psi_k \qquad \; k = 1, \ldots, n+1
\lab{3.1a}
\ee
We now see that thanks to the special form
 of $E$ in $A = \l E + A^{0}$, \rf{3.1} takes a simple and  equivalent form:
\be
\left(\begin{array}{ccccc}
\pa &0 &\cdots &0 &q_1 \\
0 & \pa &0 &\cdots &q_2 \\
\vdots & &\ddots & &\vdots \\ 0& & &\pa  & q_n\\
r_1& r_2 &\cdots &r_n & \pa+\l  \end{array} \right)
\left(\begin{array}{l}
{\bar \psi}_1 \\ {\bar \psi}_2 \\ \vdots \\ {\bar \psi}_n \\{\bar\psi}_{n+1}
\end{array} \right)   =        0
\lab{3.2}
\ee
The linear problem \rf{3.2} after elimination of ${\bar\psi}_{k}\;,\, k
= 1, \ldots, n$ takes a form of the scalar eigenvalue problem in terms
of a single eigenfunction ${\bar \psi}_{n+1}$:
\be
- \lb \pa - \sum_{k=1}^n r_k \pa^{-1} q_k \rb {\bar\psi}_{n+1}
= \l {\bar\psi}_{n+1}
\lab{3.3}
\ee
This defines the pseudo-differential operator
\be
L_n = \pa - \sum_{k=1}^n r_k \pa^{-1} q_k
\lab{3.4}
\ee
which can also be rewritten as
\be
L_n = \pa + \sum_{i=1}^n a_i \(\pa - S_i \)^{-1}
\lab{3.5}
\ee
by a simple substitution:
\be
r_i =  a_i e^{\int S_i} \qquad ; \qquad q_i  =  -e^{-\int S_i}
\quad ;\quad i = 1, \ldots , n
\lab{3.6}
\ee
The pseudo-differential  operator of the form \rf{3.4} defines
a constrained KP hierarchy \ct{oevels,konop,cheng}.
\subsection{On the Constrained KP Hierarchies}
Recall that KP flows with respect to infinite many times $t_m$
are defined by:
\be
\pa_{t_m} L = \lb \( L^m\)_+ \, , \, L \rb
\lab{3.7}
\ee
for the general KP Lax operator:
\be
L = \pa + \sum_{i=0}^{\infty} u_i \( \{ t_m \} \) \pa^{-1-i}
\lab{3.8}
\ee
The subscripts $\pm$ mean here that we only take non-negative/negative
powers of the differential operators $\pa $.

The KP hierarchy \rf{3.7} allows a straightforward
reduction to the so-called $k$-th order KdV hierarchy by
imposing the condition
\be
\( L^k\)_{-} = 0
\lab{3.9}
\ee
This condition is preserved by the flows in  \rf{3.7} due to
\be
\pa_{t_m} \( L^k\)_{-} = \lb \( L^m\)_+ \, , \, L^k \rb_{-}
= \lb \( L^m\)_+ \, , \, \( L^k\)_{-} \rb_{-}
\lab{3.10}
\ee
As noted in \ct{oevels,konop,cheng} condition \rf{3.9} can be made
less restrictive by allowing $\( L^k\)_{-} $ to be
\be
\( L^k\)_{-} = - r \pa^{-1} q
\lab{3.11}
\ee
since also this condition will be preserved by the KP flows, with
the flows of $r$ and $q$ given by:
\br
\pa_{t_m} r  &= & \( L^m\)_+ r  \lab{3.13}\\
\pa_{t_m} q  &= & - \( L^m\)_+^{*} q   \lab{3.14}
\er
where the upperscript denotes adjoint operation w.r.t. $\pa$.
In this way one can obtain as a special case
of \rf{3.11} the two-boson Lax operator:
\be
L_1 = \pa -r \pa^{-1} q = \pa + a \(\pa - S \)^{-1}
\lab{3.12}
\ee
defining a consistent reduction of KP hierarchy
with flows defined by $\pa_{t_m} L_1 = \lb \( L_1^m\)_+ \, , \, L_1 \rb$
and local bi-Hamiltonian structure \ct{BAK85,2boson}.
For the linkage of this restriction of the KP hierarchy to the additional
symmetries of the KP model see \ct{dickey}.

Obviously in expansion of $L_1 = \pa + \sum_i u_i \pa^{-1-i}$, the
coefficient $u_0$ is equal to $-rq$.
This can be understood as a symmetry  constraint \ct{oevels,konop,cheng},
which naturally generalizes to $u_0 = - \sum_{k=1}^n r_k q_k$ in the case
of many bosons.
This leads us to the Lax operator in \rf{3.4} with flows
\br
\pa_{t_m} L_n &=& \lb \( L_n^m\)_+ \, , \, L_n \rb \lab{3.15}\\
\pa_{t_m} r_i  &= & \(\(\pa - \sum_{k=1}^n r_k \pa^{-1} q_k \)^m\)_{+}r_i
 \lab{3.16}\\
\pa_{t_m} q_i  &= &- \(\(-\pa + \sum_{k=1}^n q_k \pa^{-1} r_k \)^m\)_{+} q_i
\lab{3.17}
\er
We will call the hierarchy defined by the Lax operator $L_n$ from \rf{3.4}
and flows \rf{3.15}-\rf{3.17} the n-generalized two-boson KP hierarchy.

\subsection{Equivalence Between ${\bf n}$-Generalized Two-Boson KP
Hierarchy and ${\bf sl (n+1)}$ GNLS Hierarchy}

We prove now the equivalence between $sl (n+1)$ $\GNLS$ hierarchy defined
on basis of the linear problem \rf{3.1} and
the  n-generalized two-boson KP hierarchy introduced in the previous
subsection.
Our result is contained in the following

\prop {\em Flows \rf{3.15}-\rf{3.17} of n-generalized two-boson KP hierarchy
coincide with the flows produced  by the recursion operator
 $\cR$  \rf{1.19} of the $sl (n+1)$ $\GNLS$ hierarchy }

We will generalize the proof given in \ct{cheng} for the $n=1$.
We proceed by induction.
Let us first introduce  some notation.
We parametrize the $m$-th power of $L_n$ \rf{3.5} as
\be
L_n^m = \sum_{j \leq m} P_j (m) \pa^j
\lab{3.18}
\ee
By defining
\be
B_m \equiv \( L_n^m\)_{+} = \sum_{j=0}^{m} P_j (m) \pa^j
\lab{3.19}
\ee
the flow equations \rf{3.16}-\rf{3.17} can be rewritten as
\be
\pa_{t_m} r_i  = B_m r_i \qquad;\qquad
\pa_{t_m} q_i  = - B_m^{*} q_i
\lab{3.20}
\ee
The case $m=1$ is obvious.
Let us therefore make an appropriate induction assumption about $m$,
namely that \rf{3.20} and \rf{1.19} agree. Consider now:
\br
B_{m+1} &=& \llbrack \(\pa - \sum_{k=1}^n r_k \pa^{-1} q_k \)
\sum_{j=-1}^{m} P_j (m) \pa^j \rrbrack_{+} \lab{3.21} \\
&=& \pa B_m + P_{-1} (m) + \sum_{k=1}^n \sum_{j=1}^m \sum_{i=1}^j (-1)^i
r_k \( q_k P_j (m) \)^{(i-1)} \pa^{j-i}        \nonu
\er
where $f^{(i)} \equiv \pa^i_x f$.
To calculate the constant term $P_{-1} (m)$ we note that
$P_{-1} (m)= {\rm Res} L_n^m$ which is equal to an Hamiltonian density
$\cH_m$ and therefore
\br
P_{-1} (m) \eq \cH_m = \pa^{-1} \pa_{t_m} \( -\sum_{k=1}^n r_k  q_k \)
= - \sum_{k=1}^n  \pa^{-1} \( q_k B_m r_k - r_k B_m^{*} q_k \)
\lab{3.22}\\
\eq - \sum_{k=1}^n  \sum_{j=0}^m \pa^{-1} \( q_k P_j (m) (r_k)^{(j)}
- (-1)^j r_k ( P_j (m) q_k )^{(j)} \) \nonu\\
&=& - \sum_{k=1}^n  \sum_{j=1}^m \sum_{i=1}^j (-1)^i
\( q_k P_j (m)\)^{(i-1)} (r_k)^{(j-i)} \nonu
\er
where we have used an identity
\be
- \( A B^{(j)} - (-1)^j A^{(j)} B \) =
\pa \sum_{i=1}^j (-1)^i \(A\)^{(i-1)} (B)^{(j-i)}
\lab{3.23}
\ee
valid for $j \geq 1$ and arbitrary $A$ and $B$.
With this information we can now apply $B_{m+1}$ on $r_i$:
\br
B_{m+1} r_i \eq \pa (B_m r_i) -
\sum_{k=1}^n r_i \pa^{-1} \( q_k B_m r_k - r_k B_m^{*} q_k \) \lab{3.24}
\\
&- & \sum_{k=1}^n r_k \pa^{-1} \( q_k B_m r_i - r_i B_m^{*} q_k \)
\nonu
\er
Writing now the recursion relation of the first section \rf{1.19}
with induction assumption taken into account:
\be
\pa_{m+1} \twocol{r_i}{q_l} =
\cR_{(i,l),(j,p)} \twocol{B_m r_j}{- B_m^{*} q_p}
\lab{3.25}
\ee
We find from \rf{3.25} and \rf{1.19}
\be
\pa_{m+1} r_i = - \{ \pa (B_m r_i) -
\sum_{k=1}^n r_i \pa^{-1} \( q_k B_m r_k - r_k B_m^{*} q_k \)
- \sum_{k=1}^n r_k \pa^{-1} \( q_k B_m r_i - r_i B_m^{*} q_k \) \}
\lab{3.25a}
\ee
which agrees with \rf{3.24} ( up to a total minus sign).
Similarly to \rf{3.21} we find
\be
B_{m+1}^{*} =
- \pa B_m^{*} + P_{-1} (m) + \sum_{k=1}^n \sum_{j=1}^m \sum_{i=1}^j
(-1)^{j+i} q_k (r_k)^{(i-1)} \pa^{(j-i)}  P_j (m) \lab{3.26}
\ee
Applying this on $q_l$ we get
\br
B_{m+1}^{*} q_l \eq - \pa (B_m^{*} q_l) + P_{-1} (m) q_l
+ \sum_{k=1}^n \sum_{j=1}^m \sum_{i=1}^j
(-1)^{j+i} q_k (r_k)^{(i-1)} \(  P_j (m) q_l \)^{(j-i)} \lab{3.27} \\
\eq -   \pa (B_m^{*} q_l) -
\sum_{k=1}^n q_l  \pa^{-1} \( q_k B_m r_k - r_k B_m^{*} q_k \)
-  \sum_{k=1}^n q_k \pa^{-1} \( r_k B_m^{*} q_l - q_l B_m r_k \)
\nonu
\er
following again from identity \rf{3.23}.
Again we find agreement (up to a total minus
sign) with relation \rf{1.19}  defining recursion operator
for the $sl (n+1)$ $\GNLS$ hierarchy.

\sect{Constrained KP Hierarchies from GNLS Hierarchy}
In this section we derive a class of $\CKP$ models from $\GNLS$
hierarchy using the similarity  transformations.
Let us go back to the linear problem defined by $A^{0}$ from
\rf{A1.25} or \rf{1.26}. In matrix notation it is given by \rf{4.1}.
Performing again the phase transformation \rf{3.2} and eliminating
${\bar \psi}_1, \ldots, {\bar \psi}_n$ from the matrix eigenvalue
problem we obtain a scalar eigenvalue equation:
\be
- \lb \pa + \sum_{i=1}^n (-1)^i r_i \prod_{k=i}^1 \pa^{-1} q_k \rb
\; {\bar\psi}_{n+1} = \l {\bar\psi}_{n+1}
\lab{4.2}
\ee
This linear problem defines another example of constrained KP hierarchy
involving the pseudo-differential operators given by:
\br
\cL_n &=& \pa + \sum_{i=1}^n (-1)^i r_i \prod_{k=i}^1 \pa^{-1} q_k
\lab{4.3}\\
& =& \pa + \sum_{i=1}^n a_i \(\pa - S_i \)^{-1} \cdots \(\pa - S_1 \)^{-1}
\lab{4.4}
\er
The coefficients in \rf{4.3} and \rf{4.4} are related through
\be
r_i =  a_i e^{\int S_i} \qquad ; \qquad q_i  =  - e^{\int (S_{i-1}- S_i)}
\quad ;\quad S_0 =0  \; ;\quad i = 1, \ldots , n
\lab{4.5}
\ee
The pseudo-differential operators of the type shown in \rf{4.4}
appear
naturally in connection with the Toda lattice hierarchy \ct{BX9305}.
We will refer to the corresponding hierarchy
as the multi-boson KP-Toda hierarchy.
In this section we will explain its status as a constrained KP hierarchy
and establish its connection with the Toda lattice.

In \ct{ANP93} it was shown that the first bracket
of the multi-boson KP-Toda hierarchy is a consistent reduction
of the first Poisson structure of the full KP hierarchy.
Here we will construct successive Miura maps taking the n-generalized
two-boson KP system into a sequence of constrained  KP hierarchies
ending with the multi-boson KP-Toda hierarchy.
This will establish the latter as a multi-hamiltonian reduction of the
KP hierarchy.
Our derivation will reveal the presence of discrete
Schlesinger-B\"{a}cklund symmetry of the multi-boson KP-Toda hierarchy.
We will also identify the n-generalized two-boson KP hierarchy
as an abelian structure used in the proof given in \ct{ANP93}.

\subsection{Similarity Transformations}
We start with the Lax operator of the n-generalized two-boson
KP hierarchy \rf{3.5} and define a new variable
$S_1^{(1)} \equiv S_1 + \pa \ln a_1$.
Next we perform the following similarity transformation:
\br
L_n^{(1)} \! &\equiv&\! \( \pa - S^{(1)}_1 \) \; L_n \;
\( \pa - S^{(1)}_1 \)^{-1} \nonu \\
&=&\pa + a^{(1)}_1 \( \pa - S^{(1)}_1 \)^{-1}  + \sum_{i=2}^n a^{(1)}_i
\( \pa - S_i \)^{-1} \( \pa - S^{(1)}_1 \)^{-1}
\lab{4.6}
\er
where we have introduced the redefined coefficients:
\br
a^{(1)}_1 \! &\equiv&\! {S^{(1)}_1}^{\pr} + \sum_{i=1}^n a_i \lab{4.7}\\
a^{(1)}_i  \! &\equiv&\! a^{\pr}_i +  a_i \( S_i - S^{(1)}_1 \)
\qquad i=2, \ldots,n
\lab{4.8}
\er
If $a_i =0 $ for $ i=3, \ldots,n$ we would
have already obtained in this way the Lax
operator  of the four-boson KP-Toda hierarchy
{}from \rf{4.4}.
Otherwise we have to continue to apply successively the similarity
transformations. In such case, the next step is:
\br
L_n^{(2)} \! &\equiv&\! \( \pa - S^{(2)}_2 \) \; L^{(1)}_n \;
\( \pa - S^{(2)}_2 \)^{-1} =\pa + a^{(2)}_1 \( \pa - S^{(2)}_2 \)^{-1}
+ a^{(2)}_2 \( \pa - S^{(1)}_1 \)^{-1} \( \pa - S^{(2)}_2 \)^{-1}
\nonu \\
& +& \sum_{i=3}^n a^{(2)}_i
\( \pa - S_i \)^{-1} \( \pa - S^{(1)}_1 \)^{-1}\( \pa - S^{(2)}_2 \)^{-1}
\lab{4.9} \\
S_2^{(2)} &\equiv& S_2 + \pa \ln a_2^{(1)} \lab{4.10}\\
a^{(2)}_1 \! &\equiv&\! a_1^{(1)} + {S_2^{(2)}}^{\pr} \lab{4.11} \\
a^{(2)}_2 \! &\equiv&\! a_2^{(1)} + \( \pa + S^{(1)}_1 - S_2^{(2)} \)
a^{(1)}_1  + \sum_{i=3}^n a_i^{(1)} \lab{4.12}\\
a^{(2)}_i  \! &\equiv&\!  \( \pa + S_i - S^{(2)}_2 \) a_i^{(1)} \qquad
3 \geq i \geq n \lab{4.13}
\er
This defines a string of successive similarity transformations.
The next step will involve similarity transformation
$\( \pa - S^{(3)}_3 \) \; L^{(2)}_n \;
\( \pa - S^{(3)}_3 \)^{-1}$ with $S_3^{(3)} \equiv S_3 + \pa \ln a_3^{(2)}$
and so on.
After $k$ steps we arrive at:
\br
&&L_n^{(k)} \equiv \pa + a^{(k)}_1 \( \pa - S^{(k)}_k \)^{-1}
+ a^{(k)}_2 \( \pa - S^{(k-1)}_{k-1} \)^{-1} \( \pa - S^{(k)}_k \)^{-1}
+ \ldots \lab{4.14} \\
&&+ a^{(k)}_k \( \pa - S^{(1)}_{1} \)^{-1} \cdots
\( \pa - S^{(k)}_k \)^{-1} + \sum_{i=k+1}^n a^{(k)}_i
\( \pa - S_i \)^{-1} \( \pa - S^{(1)}_1 \)^{-1} \cdots
\( \pa - S^{(k)}_k \)^{-1}
\nonu \\
&& S_k^{(k)} \equiv S_k + \pa \ln a_k^{(k-1)} \lab{4.15}\\
&& a^{(k)}_1 \equiv  a_1^{(k-1)} + {S_k^{(k)}}^{\pr} \lab{4.16} \\
&& a^{(k)}_l \equiv a_l^{(k-1)} + \( \pa + S^{(k-l+1)}_{k-l+1}
- S_k^{(k)} \) a^{(k-1)}_{l-1} \qquad l=2, \ldots, k-1 \lab{4.17} \\
&& a^{(k)}_k \equiv a_k^{(k-1)} + \( \pa + S^{(1)}_{1}
- S_k^{(k)} \) a^{(k-1)}_{k-1}
 + \sum_{p=k+1}^n a_p^{(k-1)} \lab{4.18}\\
&& a^{(k)}_p  \equiv  \( \pa + S_p - S^{(k)}_k \) a_p^{(k-1)} \qquad
k+1 \geq p \geq n \lab{4.19}
\er
If $ n$ is equal to $k+1$ we have obtained in equation \rf{4.14}
the Lax operator $L_n^{(n-1)}$ of the form given in \rf{4.4} and
therefore member of the n-boson KP-Toda hierarchy.
Since the similarity transformations do not change Hamiltonians, the new
hierarchy of the Lax operators from \rf{4.4} will share
 the infinite set of
involutive Hamiltonians with the n-generalized two-boson KP hierarchy.
The corresponding Poisson bracket structures are obtained by applying Miura
maps
defined by \rf{4.15}-\rf{4.19}.

At this point we would like to remark on the following ambiguity connected with
our formalism of successive similarity transformations.
Note, that $L^{(n-1)}_n $ obtained from \rf{4.14} by setting $k=n-1$
can be further transformed by an extra similarity transformation without
changing its form.
This is achieved by:
\br
L_n^{(n)} \! &\equiv&\! \( \pa - S^{(n)}_n \) \; L^{(n-1)}_n \;
\( \pa - S^{(n)}_n \)^{-1} \nonu \\
&=& \pa + \sum_{l=1}^n a^{(n)}_l
\( \pa - S^{(n-l+1)}_{n-l+1} \)^{-1} \cdots \( \pa - S^{(n)}_n \)^{-1}
\lab{4.20}
\er
where
\br
S_n^{(n)} &\equiv& S_n + \pa \ln a_n^{(n-1)} \lab{4.21}\\
a^{(n)}_1 \! &\equiv&\! a_1^{(n-1)} + {S_n^{(n)}}^{\pr} \lab{4.22} \\
a^{(n)}_l \! &\equiv&\! a_l^{(n-1)} + \( \pa + S^{(n-l+1)}_{n-l+1}
- S_n^{(n)} \) a^{(n-1)}_{l-1} \qquad l=2, \ldots, n \lab{4.23}
\er
In \rf{4.21}-\rf{4.22} we recognize the Toda lattice structure.
Hence the ambiguity encountered in associating
the multi-boson KP-Toda hierarchy Lax operator to the underlying n-generalized
two-boson KP hierarchy is an origin of the discrete symmetry of the
multi-boson KP-Toda hierarchy \ct{similar}.  The discrete symmetry is in
this context the similarity map $L^{(n-1)} \rightarrow L^{(n)}$.

\subsection{Eigenfunctions  and Flow Equations
for the Constrained KP Hierarchies}
Let us start with a technical observation that the similarity
transformation, which takes $L_n^{(k-1)}$ to $L_n^{(k)} $ :
\be
L_n^{(k)} =  \( \pa - S^{(k)}_k \) \; L^{(k-1)}_n \;
\( \pa - S^{(k)}_k \)^{-1}
\lab{simex}
\ee
can be equivalently written as
\be
L_n^{(k)} =  \Phi^{(k-1)}_k \pa {\Phi^{(k-1)}_k}^{-1} \; L^{(k-1)}_n \;
\Phi^{(k-1)}_k \pa^{-1} {\Phi^{(k-1)}_k}^{-1}
\lab{simgauge}
\ee
where we have defined:
\be
\Phi^{(k-1)}_{k} \equiv e^{ \int S^{(k)}_{k}} =
a^{(k-1)}_{k} e^{\int S_{k}}
\lab{sphidef}
\ee
This allows us to find a compact expression for a string of successive
similarity transformations leading from $L_n$ to $L_n^{(k)} $:
\be
L_n^{(k)} = \prod_{l=k-1}^0 \( \Phi^{(l)}_{l+1} \pa {\Phi^{(l)}_{l+1}}^{-1}
\) \; L_n \;
 \prod^{k-1}_{l=0} \( \Phi^{(l)}_{l+1} \pa^{-1} {\Phi^{(l)}_{l+1}}^{-1} \)
\lab{compsim}
\ee

It is convenient at this point to introduce a definition of the eigenfunction
for the Lax operator $L$.

\name {\em A function $\Phi$ is called eigenfunction for the Lax
 operator $L$ satisfying
Sato's flow equation \rf{3.7} if its  flows are given by
expression:}
\be
\pa_{t_m} \Phi = \( L^m\)_{+} \Phi
\lab{eigenlax}
\ee
{\em for the infinite many times $t_m$}

In particular as we have seen in Section 3 the Lax $L_n$ \rf{3.5}
of the n-generalized two-boson KP hierarchy possesses $n$ eigenfunctions
$\Phi^{(0)}_n \equiv r_i = a_i \exp \({\int S_i}\)$.

We will now establish a main result of this subsection
for the Lax $L^{(k)}_n$ defined in \rf{4.14}.

\prop {\em The Lax operators $L^{(k)}_n$ \rf{4.14} satisfy Sato's
hierarchy equations \rf{3.7} and possess $n-k$ eigenfunctions given by: }
\be
\Phi^{(k)}_p = a^{(k)}_p e^{\int S_p} \qquad p = k+1 , \ldots, n
\lab{eigenkp}
\ee
Especially we find that $\Phi^{(k)}_{k+1} = \exp \( \int S^{(k+1)}_{k+1}
\)$.
We will base our induction proof on the result  of \ct{oevelr}
that for given two eigenfunctions $\Phi_1$, $\Phi_2$
of the Lax operator $L$ satisfying Sato's hierarchy equation \rf{3.7}
the Lax ${\hat L} \equiv \Phi_1 \pa {\Phi_1}^{-1} \, L \,
\Phi_1 \pa^{-1} {\Phi_1}^{-1} $ also satisfies the
hierarchy equation \rf{3.7}
and the function $\Phi_1 \(\Phi_1^{-1}\Phi_2\)^{\pr}$
is an eigenfunction of ${\hat L}$.

Start with $k=1$. Define $\Phi^{(1)}_i \equiv r_1
\( r_1^{-1} r_i\)^{\pr} = \( \pa - (\pa \ln r_1)\) r_i$ for
$i=2,\ldots, n $.
{}From the result of \ct{oevelr} stated in the previous paragraph and
equation \rf{simgauge} for $k=1$ we find that
$\Phi^{(1)}_i$ is an eigenfunction of hierarchy of $L^{(1)}_n$.
Furthermore substituting $r_i$'s by $a_i$ and $S_i$ according to
\rf{4.5} we find the desired result
$\Phi^{(1)}_i= a_i^{(1)} \exp \({\int S_i}\)$.

Let us now assume that $\Phi^{(k)}_p = a^{(k)}_p e^{\int S_p}$
for $p = k+1 , \ldots, n$ are indeed eigenfunctions belonging to the
flow hierarchy of $L^{(k)}_n$.
Define now
\br
\Phi^{(k+1)}_p &\equiv& \( \pa - (\pa \ln \Phi^{(k)}_{k+1} \)
\Phi^{(k)}_{p} = \( \pa - (\pa \ln a^{(k)}_{k+1}) - S_{i+1}\)
a^{(k)}_p e^{\int S_p} \lab{indu1}\\
&=& \lb \( \pa - (\pa \ln a^{(k)}_{k+1}) - S_{i+1} +S_p\)
a^{(k)}_p \rb e^{\int S_p}=  a^{(k+1)}_p e^{\int S_p}
 \lab{indu2}
\er
where in \rf{indu2} we have used \rf{4.19}.
Since
\be
L^{(k+1)}_n
 =  \Phi^{(k)}_{k+1} \pa {\Phi^{(k)}_{k+1}}^{-1} \; L^{(k)}_n \;
\Phi^{(k)}_{k+1} \pa^{-1} {\Phi^{(k)}_{k+1}}^{-1}
\lab{since}
\ee
it follows from induction assumption that $L^{(k+1)}_n $ satisfies
\rf{3.7}. Furthermore since
expression in \rf{indu1} is equal to
$\Phi^{(k+1)}_p \equiv \Phi^{(k)}_{k+1} \({\Phi^{(k)}_{k+1}}^{-1}
\Phi^{(k)}_{p}\)^{\pr}$ we have found eigenfunctions belonging
to $L^{(k+1)}_n$ and they are given by \rf{eigenkp}.
This concludes the induction proof.

As a corollary we find that the Lax operators $\cL_n= L^{(n)}_n$
of the multi-boson KP-Toda hierarchy satisfy Sato's hierarchy equations
\rf{3.7}.

Successive gauge transformations of the type given in \rf{simgauge}
resulting in a Lax structures with decreasing number of eigenvalues
were also considered in \ct{oevela}.

\subsection{On the Discrete Schlesinger-B\"{a}cklund Transformation
of the Generalized Toda-AKNS model}
In this subsection we will study the discrete B\"{a}cklund
transformations obtained above in the Lax formulation within the
corresponding matrix formulation of $\GNLS$ hierarchy.

To set the scene let us first connect a ``two-boson"
Toda system with AKNS matrix formulation.
 The ``two-boson" Toda spectral system is:
\br
\( \pa - a_0 (n-1)\) \Psi_{n-1} \eq \Psi_n     \lab{speca}\\
 \Psi_{n+1}  + a_0 (n) \Psi_n + a_1 (n) \Psi_{n-1} \eq \l \Psi_n
\lab{specb}
\er
which becomes in matrix formulation:
\be
\pa \twocol{\Psi_{n-1}}{\Psi_{n}}
= \fourmat{a_0 (n-1)}{1}{-a_1 (n)}{\l} \twocol{\Psi_{n-1}}{\Psi_{n}}
= A_n (t, \l ) \twocol{\Psi_{n-1}}{\Psi_{n}}
\lab{mattoda}
\ee
Under the lattice shift $ n \to n+1$, $\Psi_n$ goes
\be
\Psi_{n+1} = \l \Psi_n - a_0 (n) \Psi_n - a_1 (n) \Psi_{n-1}
= \(\l - a_0 (n)\) \Psi_n - a_1 (n) \Psi_{n-1}
\lab{nplusone}
\ee
Recalling that $a_0 (n) = a_0 (n-1) + \pa \ln a_1 (n) $
we can describe the transition $n \to {n+1}$ in form of
the matrix equation involving only variables entering \rf{mattoda}:
\be
\twocol{\Psi_{n}}{\Psi_{n+1}}=
\fourmat{0}{1}{-a_1 (n)}{\(\l-a_0 (n-1) - \pa \ln a_1 (n)\)}
\twocol{\Psi_{n-1}}{\Psi_{n}} \equiv T_n (t, \l)
\twocol{\Psi_{n-1}}{\Psi_{n}}
\lab{fnplusone}
\ee
Compatibility of \rf{mattoda} and \rf{fnplusone} yields:
\be
A_{n+1} = T_n A_{n}T_n^{-1}+T_n \pa T_n^{-1}
\lab{gaugesl}
\ee
We will now establish the relation to the AKNS hierarchy.
Introduce the following new variables:
\be
\twocol{\p^{(2)}_{n}}{\p^{(1)}_{n}}
= \twocol{e^{-\int a_0 (n-1)} \Psi_{n-1}}{\Psi_{n}}
\lab{aknsp}
\ee
and
\be
q_n \equiv -e^{-\int a_0 (n-1)} \qquad;\qquad
r_n \equiv a_1 (n) e^{\int a_0 (n-1)}
\lab{rqdef}
\ee
Now we can rewrite \rf{speca}-\rf{specb} in matrix notation as
\be
\fourmat{\pa}{q_n}{r_n}{\pa - \l} \twocol{\p^{(2)}_{n}}{\p^{(1)}_{n}}
= 0         \lab{almost}
\ee
If we let $\p^{(i)}_{n} \to e^{-\int^x \l /2} \p^{(i)}_{n}$ we arrive at
exactly AKNS equation:
\be
\fourmat{\pa+ \h \l}{q_n}{r_n}{\pa -\h \l}
\twocol{e^{-\int^x \l /2}\p^{(2)}_{n}}{e^{-\int^x \l /2}\p^{(1)}_{n}} = 0
  \lab{aknseq}
\ee
Using the Toda chain equations
\br
\pa a_0 (n) \eq a_1 (n+1) - a_1 (n)  \lab{t1eqa} \\
\pa a_1 (n) \eq  a_1 (n) \( a_0 (n) - a_0 (n-1) \)
           \lab{t1eqb}
\er
and \rf{nplusone} or \rf{fnplusone} we find that the shift $n \to n-1$
of the Toda lattice results in
\be
\fourmat{\l- \pa \ln q_n}{q_n}{-1 /q_n}{0}
 \twocol{\p^{(2)}_{n}}{\p^{(1)}_{n}} =
  \twocol{\p^{(2)}_{n-1}}{\p^{(1)}_{n-1}}
\lab{nnminusq}
\ee
together with
\be
q_n \to q_{n-1} = - r_n q_n^2 + \pa^2 q_n - { (\pa q_n)^2 \o q_n}
\quad;\quad   r_n \to r_{n-1} = - {1 \o q_n}
\lab{qrnminus}
\ee
while the shift $n \to n+1$ of the Toda lattice
results in (equivalently to \rf{fnplusone})
\be
\fourmat{0}{r^{-1}_n}{-r_n}{(\l - \pa \ln r_n)}
 \twocol{\p^{(2)}_{n}}{\p^{(1)}_{n}} =
  \twocol{\p^{(2)}_{n+1}}{\p^{(1)}_{n+1}}
\lab{nnplus}
\ee
together with
\be
q_n \to q_{n+1} = - {1 \o r_n}
\quad;\quad   r_n \to r_{n+1} = - r_n^2 q_n +
\pa^2 r_n - { (\pa r_n)^2 \o r_n}
\lab{qrnplus}
\ee
These are so-called Schlesinger transformations  discussed in
\ct{fla83,ne85,BK88}.
Here we obtained them from a lattice shift of the Toda lattice underlying
the AKNS construction.
The similarity transformation on the level of corresponding
Lax operator captures, as shown in a previous subsection, a Toda lattice
structure within the continuous constrained KP hierarchy and is fully
equivalent to the discrete Schlesinger-B\"{a}cklund transformation
obtained here  because \rf{4.21} and \rf{4.23} correspond to the Toda
chain equations \rf{t1eqa} and \rf{t1eqb}.

For completeness let us comment on the Schlesinger-B\"{a}cklund
transformations in the case of the general Toda hierarchy.
We start this time with the spectral equation:
\br
\pa \Psi_n \eq \Psi_{n+1} + a_0 (n) \Psi_n     \nonu \\
\l \Psi_n \eq \Psi_{n+1}  + a_0 (n) \Psi_n
+ \sum_{k=1}^M a_k (n) \Psi_{n-k}  \lab{spectr}
\er
The spectral equation \rf{spectr} can be rewritten as a matrix equation,
which in components is given by:
\be
\left(\begin{array}{ccccc} \pa- a_0 (n-M) & -1 &0 & \ldots &0 \\
0& \pa- a_0 (n-M+1) & -1 &0 & \ldots \\
\vdots & \vdots &\ddots &\ddots &\vdots \\
0& 0&0  &\pa- a_0 (n-1)  & -1  \\
a_M (n) & a_{M-1} (n)& \ldots & a_1 (n) &\pa - \l \end{array} \right)
\left(\begin{array}{c}
\Psi_{n-M} \\ \Psi_{n-M+1} \\ \vdots \\ \Psi_{n-1} \\
\Psi_{n} \end{array} \right)  = 0
\lab{matrix}
\ee
Under the transition $n \to n+1$ on the lattice we get:
\be
\left(\begin{array}{l}
\Psi_{n-M+1} \\ \Psi_{n-M+2} \\ \vdots \\  \Psi_{n} \\ \Psi_{n+1}
\end{array} \right)  =
\left(\begin{array}{ccccc}
0 &1 &0 &\cdots &0 \\
0 & 0 &1 &\cdots &0 \\
\vdots & &\vdots &\ddots &0 \\ 0&\cdots &0 &\ddots  & 1\\
-a_M (n)& -a_{M-1}(n) &\cdots &- a_1 (n) & (\l - a_0 (n) )
\end{array} \right)
\left(\begin{array}{l}
\Psi_{n-M} \\ \Psi_{n-M+1} \\ \vdots \\  \Psi_{n-1} \\ \Psi_{n}
\end{array} \right)
\lab{5.3}
\ee
Introduce now AKNS notation:
\br
r_i (n) &=& a_i (n) e^{\int a_0 (n-i)} \quad ; \quad q_i = -
e^{\int (a_0 (n-i) - a_0 (n-i-1)} \quad\; i= 1,\ldots,M
\lab{5.1}\\
\p^{(1)}_n &\equiv& \Psi_n \quad ;\quad
\p^{(i)}_n \equiv e^{-\int a_0 (n-i+1)}\Psi_{n-i+1} \quad ;\quad
i = 2, \ldots , M+1
\lab{5.2}
\er
where for consistency we set $a_0 (n-M-1)=0$.
The relation \rf{5.3} reads
in the AKNS variables as:
\be
\left(\begin{array}{l}
\p^{(M+1)}_{n+1} \\ \p^{(M)}_{n+1} \\ \vdots \\  \p^{(2)}_{n+1} \\
\p^{(1)}_{n+1} \end{array} \right)
=\left(\begin{array}{ccccc}
0 &1 &0 &\cdots &0 \\
0 & 0 &1 &\cdots &0 \\
\vdots & &\vdots &\ddots &0 \\ 0&\cdots &0 &\ddots  & r_1^{-1} (n)\\
-r_M (n)& -r_{M-1} (n) &\cdots &- r_1 (n) & \(\l - \pa \ln r_1 (n) \)
  \end{array} \right)
\left(\begin{array}{l}
\p^{(M+1)}_{n} \\ \p^{(M)}_{n} \\ \vdots \\  \p^{(2)}_{n} \\
\p^{(1)}_{n}
\end{array} \right)
\lab{mschles}
\ee
This transformation generalizes the Schlesinger-B\"{a}cklund
transformation from the AKNS system to a general AKNS system based on
arbitrary Toda lattice.
\appendix
\sect{Lie Algebra Preliminaries}
In order to be self contained we recall some  basic results on the theory
of Lie  algebras. We first establish the commutation relations in the
Chevalley basis,
 \br
\{ H_a\, ,\, H_b\} & = & 0 \nonumber \\
\{ H_a \, ,\, E_{\a } \} & = & K_{\a a} E_{\a } \nonumber
\er
\be
\{ E_{\a }\, , \,  E_{\b} \}  =
\left\{ \begin{array}{cl}
 \eps (\a ,\b ) E_{\a +\b} & \a +\b\quad {\rm is~a~root} \\
 \sum _{a=1}^{n} n_a H_a & \a +\b =0 \\
 0 &{\rm otherwise} \end{array} \right.
\lab{sunalg}
\ee
where $K_{\a a} = {2\a . \a_a \o{\a_a^2}} = \sum n_bK_{ba}$, for $K_{ab}$
the Cartan matrix.  A root $\a$ can be expanded in terms of simple roots
as $\a = \sum n_a \a_a$.  The integers $l_a$ are defined from the
expansion ${\a \o {\a^2}} = {\sum l_a \a_a \o {\a_a^2}}$.
$\eps (\a, \b)$ constants
are related among themselves by the Jacobi identities and the antisymmetry
of the bracket.  In particular they satisfy
\be
\eps (\a,\b) = - \eps (\b,\a) = \eps(-\b,-\a)
\lab{A2}
\ee
and
\be
\eps (\a,\b) = \eps (\b,\g) = \eps (\g,\a)
\lab{A3}
\ee
for $\a +\b+\g = 0$.

We choose a special element $E$ in the Lie algebra defined as
in terms of fundamental weights
$\mu_a$, where $2\mu_a .\a_b /{\a_b^2} = \d_{ab}$ as
\be
E={2\mu_a .H \o{\a_a^2}}
\lab{A4}
\ee
The element E decomposes the Lie algebra
 $\lie = \cM + \cK = Ker(ad E) + Im(ad E)$
where $\cK$ is composed of all generators of $\lie$ commuting with $E$ and
is spanned by $\{ H_a, E_{\a}\}$ with $\a$ not containing the simple root
$\a_a$ while $\cM$ is its orthogonal complement.  Hermitean symmetric
spaces are associated to those Lie algebras $\lie$ with roots such
 that they either do not contain $\a_a$ or contain it only once, i.e.
\be
{2\mu_a .\a \o{\a_a^2}} = \pm 1 ,0
\lab{A5}
\ee
for all roots of $\lie$. This fact implies that
\be
\lb E \, ,\lb E\, ,\, E_{\a}\rb \rb  = E_{\a} \quad\;\rm{or} \quad\; 0
\lab{A6}
\ee
{}From the Jacobi identities it can be shown that (see e.g. \ct{crumey})
$\lie / \cK$
is a symmetric space,i.e.
\be
\lb \cK\, ,\, \cK \rb  \in \cK,\quad \lb \cM \, , \, \cK \rb \in \cM, \quad
\lb \cM \, , \, \cM \rb \in \cK,
\lab{A7}
\ee
The Lie algebras satisfying \rf{A5} are $su(n), so(2n), sp(n), E_6,and
E_7$ and generate the following Hermitian symmetric spaces
${su(n+1)\o{su(n)\times u(1)}}, \,\, {so(2n)\o u(n)}, \,\, {sp(n)\o u(n)},
\,\, {E_6 \o so(10)\times u(1)} \,\, and \,\,{E_7 \o E_6 \times u(1)}$.

The curvature tensor associated to these symmetric spaces is defined as
\be
R^{\d}_{\a\b-\g}E_{\d} = [E_{\a},[E_{\b},E_{-\g}]
\lab{A8}
\ee
with the hermiticity property
\be
R^{\d\,*}_{\a\b-\g} = R^{-\d}_{-\a \, -\b \, \g}
\lab{A8a}
\ee
\sect{${\bf sl (3)}$ ZS-AKNS Matrix Model Solution for (1.3)}
Recursion equations for Z-S approach for the case  of $A^{(0)}$
as in \rf{1.26} for $n=2$.
In this case \rf{1.32} gives:
\be
Y_m =- \pa^{-1} \pa_{m-1} q_1 E_{\a_1} = \left(\begin{array}{ccc}
{0} & - \pa^{-1} \pa_{m-1} q_1 & {0} \\ {0} & {0} & {0} \\
{0} & {0} & {0} \end{array} \right)
\lab{B1}
\ee
while \rf{1.33} yields
\be
O_m^{\cM} (1) = O_{\a_2} E_{\a_2} + O_{-\a_2} E_{-\a_2} +
O_{\a_1+\a_2} E_{\a_1+\a_2} + O_{-\a_1-\a_2} E_{-\a_1-\a_2}
\lab{B2}
\ee
with the coefficients:
\br
O_{\a_2} \eq \pa_{m-1} q_2 \qquad;\qquad O_{-\a_2} = - \pa_{m-1} r_2
+ r_1 \pa^{-1}\pa_{m-1} q_1       \lab{B3}\\
O_{\a_1+\a_2} \eq q_2 \pa^{-1}\pa_{m-1} q_1 \qquad;\qquad
O_{-\a_1-\a_2} = - \pa_{m-1} r_1
\lab{B4}
\er
plugging this into algebraic parts of \rf{1.30}-\rf{1.31} we obtain
expression for
\be
O_m^{\cK} (1) = O_{\a_1} E_{\a_1} + O_{-\a_1} E_{-\a_1} +
O_{h_1} H_{1} + O_{h_2} H_{2}
\lab{B5}
\ee
with
\br
O_{\a_1} \eq {1 \o q_2} \( \pa  q_2 \pa^{-1} \pa_{m-1} q_1  +
q_1 \pa_{m-1} q_2\)  \lab{B6} \\
O_{-\a_1} \eq \pa^{-1} \pa_{m-1} r_1 q_2  \lab{B7} \\
O_{h_1} \eq \pa^{-1} \( q_1 \pa^{-1} \pa_{m-1} (q_2 r_1 )- r_1 q_2
\pa^{-1} \pa_{m-1} q_1\)  \lab{B8} \\
O_{h_2} \eq - \pa^{-1} \pa_{m-1} q_2 r_2 \lab{B9}
\er
After inserting back into the dynamical parts of \rf{1.30}-\rf{1.31}
the above results lead to the recursion operator of the same form as the
one for the four-boson KP-Toda hierarchy.
\lskip
{\bf Acknowledgements.} H.A. is grateful to IFT-UNESP for the kind
hospitality and FAPESP for financial support.
We are indebted to M. J. Bergvelt, L. Dickey, E. Nissimov and S. Pacheva
for correspondence on the subject of this paper.

\end{document}